# Maxwell's Displacement Current and the Magnetic Field between Capacitor Electrodes

Toshio Hyodo

Slow Positron Facility, Institute of Materials Structure Science,

High Energy Accelerator Research Organization (KEK)

1-1 Oho, Tsukuba, Ibaraki, Japan 305-0801

**Abstract**

A long-standing controversy concerning the causes of the magnetic field in and around a parallel-plate capacitor is examined. Three possible sources of contention are noted and detailed. The first is the ambiguous initial impression given by the calculation of the magnetic field using the integral form of the Ampere-Maxwell law which incorporates the displacement current density. The second is misinterpretation of this law as a cause-effect formula. The third is insufficient recognition of the fact that the electric field in Maxwell's equations represents the sum of the well-distinguished irrotational and divergence-free fields, which are independently responsible for conservation of charge and the existence of the electromagnetic waves, respectively.

## 1. Introduction

The displacement current density introduced by Maxwell in his theory of electromagnetism has long been a topic of debate. (Although the concept of the electric displacement $\boldsymbol{D}(\boldsymbol{r},t)$ already carries a notion of *surface density*, here for clarity we call $\partial \boldsymbol{D}(\boldsymbol{r},t)/\partial t$ the displacement current density and its surface integral the displacement current.) A typical case of contention is whether the magnetic field in and around the space between the electrodes of a parallel-plate capacitor is created by the displacement current density in the space. History of the controversy was summarized by Roche [1], with arguments that followed [2-4] showing the subtlety of the issue. Notwithstanding that Maxwell himself thought that the displacement current gave rise to the magnetic field (just as the conduction current did) [5], it was later verified not to be the case for the part of the displacement current density that is proportional to the time derivative of the conservative (irrotational or longitudinal) electric field related to the scalar potential [6-10]. It would be worth noting that the points of contention in the apparently complicated arguments [1-4] between Roche and Jackson were substantially on the details, subject to agreement on the main point that $\partial \boldsymbol{D}/\partial t$ between capacitor electrodes is not the true source of the magnetic field.

However this characteristic of the displacement current density has not been widely recognized yet compared with the following two: (i) that it makes Maxwell's equations compatible with the conservation of charge; and (ii) that it plays a crucial role in deriving the differential equation for the electromagnetic wave. In the present article I revisit this topic and explore the issue in some detail to clarify perception and to prevent confusion. It is hoped that the discussion is accessible to undergraduate students with the aim of engendering a clear understanding of this point from when first encountering the notion of displacement current density. Thus I avoid discussions requiring the use of the electromagnetic (scalar and vector) potentials.

The following section introduces the displacement current density and the application of the resultant Ampere-Maxwell law to calculate magnetic field (commonly found in textbooks and lectures). In Section 3, use of the Biot-Savart law and Ampere-Maxwell law for the calculation of the magnetic field is discussed. The experiments claiming to have observed the displacement current are briefly reviewed in Section 4. In Section 5, Maxwell's equations are carefully deciphered from the reference point that the electric field in the equations is the sum of the conservative field and the electromagnetic induction field, with the aim of clearing up misconceptions about the role of displacement current density. A summary is given in Section 6.

## 2   Calculation of the magnetic field using the integral form of the Ampere-Maxwell law

Before we proceed, let us summarize Maxwell's equations. The electric field $\boldsymbol{E}(\boldsymbol{r},t)$ and the magnetic field $\boldsymbol{B}(\boldsymbol{r},t)$ at a position $\boldsymbol{r}$ and time $t$ in an inertial frame of reference is defined by the Lorentz force $\boldsymbol{F}$ on a charge $q$ moving at a velocity $\boldsymbol{v}$

$$\boldsymbol{F} = q(\boldsymbol{E}(\boldsymbol{r},t) + \boldsymbol{v} \times \boldsymbol{B}(\boldsymbol{r},t)) \qquad (1)$$

This means that $\boldsymbol{E}(\boldsymbol{r},t)$ is defined from the force on a charge $q$ regardless of its velocity and $\boldsymbol{B}(\boldsymbol{r},t)$ is defined from the force on a moving charge only.

The other relevant fields, the electric displacement (electric flux density) $\boldsymbol{D}(\boldsymbol{r},t)$ and the magnetic field intensity $\boldsymbol{H}(\boldsymbol{r},t)$, are related with the sources of the fields. The relationships of these fields in free space are

$$\boldsymbol{D}(\boldsymbol{r},t) = \varepsilon_0 \, \boldsymbol{E}(\boldsymbol{r},t) \qquad (2)$$

$$\boldsymbol{H}(\boldsymbol{r},t) = \mu_0^{-1} \, \boldsymbol{B}(\boldsymbol{r},t) \qquad (3)$$

where $\varepsilon_0$ is the permittivity of free space (electric constant) and $\mu_0$ is the permeability of free space (magnetic constant).

By using these four fields together with charge density $\rho(\boldsymbol{r},t)$ and current density $\boldsymbol{j}(\boldsymbol{r},t)$, Maxwell's equations, in both integral and differential forms, are written as follows:

$$\oint_S \boldsymbol{D}\cdot d\boldsymbol{S} = \int_V \rho\, dV, \qquad \nabla\cdot\boldsymbol{D} = \rho \qquad (4)$$

$$\oint_C \boldsymbol{E}\cdot d\boldsymbol{r} = -\frac{d}{dt}\int_S \boldsymbol{B}\cdot d\boldsymbol{S}, \qquad \nabla\times\boldsymbol{E} = -\frac{\partial \boldsymbol{B}}{\partial t} \qquad (5)$$

$$\oint_S \boldsymbol{B}\cdot d\boldsymbol{S} = 0, \qquad \nabla\cdot\boldsymbol{B} = 0 \qquad (6)$$

$$\oint_C \boldsymbol{H}\cdot d\boldsymbol{r} = \int_S \left(\boldsymbol{j}+\frac{\partial \boldsymbol{D}}{\partial t}\right)\cdot d\boldsymbol{S}, \qquad \nabla\times\boldsymbol{H} = \boldsymbol{j}+\frac{\partial \boldsymbol{D}}{\partial t} \qquad (7)$$

where, S in (4) and (6) indicates the closed surface enclosing an arbitrary volume V, and S in (5) and (7) indicates arbitrary open surface bounded by an arbitrary closed path C. In (5) and (7) $d\boldsymbol{S}(=\boldsymbol{n}dS)$ is the surface element having the normal $\boldsymbol{n}$ directed in accordance with the right-hand rule with the direction of $d\boldsymbol{r}$ in the circulation integral along C from the left-hand side of the equations. Note, I have dropped the notation $(\boldsymbol{r},t)$ in the equations that hold at any position and time in a frame of reference.

It is well known that Maxwell added the displacement current density term $\partial \boldsymbol{D}/\partial t$ to Ampere's law,

$$\oint_C \boldsymbol{H}\cdot d\boldsymbol{r} = \int_S \boldsymbol{j}\cdot d\boldsymbol{S}, \qquad \nabla\times\boldsymbol{H} = \boldsymbol{j} \qquad (8)$$

obtaining (7), now called the Ampere-Maxwell law, to make it work for the case of an open circuit containing a capacitor as shown in figure 1. It is assumed that the resistance of the leads and the electrodes is negligible.

If the circuit is a long straight line without the capacitor and with a current $I$ flowing, one may apply (8) to find the magnetic field at point $P_1$, distance $R$ away from the current. Applying the integral form of the law to a circle $C_1$ centered on the current and through the point $P_1$ together with flat surface $S_1$ perpendicular to the current and bordered by $C_1$, one gets

$$\oint_{C_1} \boldsymbol{H} \cdot d\boldsymbol{r} = \int_{S_1} \boldsymbol{j} \cdot d\boldsymbol{S} = I \qquad (9)$$

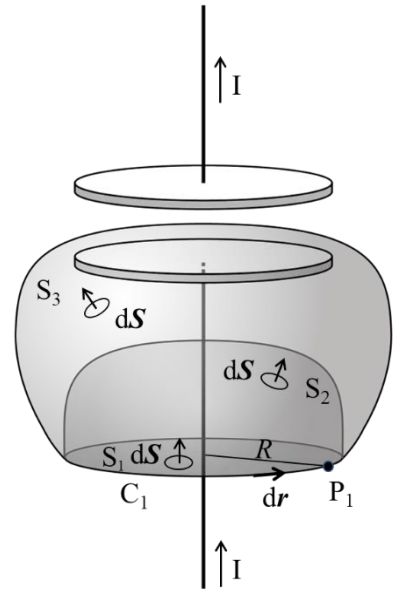

Figure 1. A circular parallel-plate capacitor being charged by the current $I$ in long straight wires. A circle $C_1$ of radius $R$ and surfaces $S_1$-$S_3$ bordered by $C_1$ are used to calculate the magnetic field at point $P_1$ on $C_1$. The surface element vectors $d\boldsymbol{S}$ for the surfaces $S_1$-$S_3$ are also shown. Their directions follow the right-hand rule with the direction of the line element $d\boldsymbol{r}$ for the circle $C_1$

Noting the axial symmetry, one gets the magnitude of the magnetic field intensity at $P_1$ as

$$H_1 = \frac{I}{2\pi R} \qquad (10)$$

In the case shown in figure 1 where we have the capacitor, if $P_1$ is located far away from the capacitor, $\boldsymbol{H}$ has the azimuthal component alone and so the field is calculated in the same way to obtain (10). Use of $S_2$ instead of $S_1$ gives the same result. However, if one chooses the surface $S_3$,

$$\oint_{C_1} \boldsymbol{H} \cdot d\boldsymbol{r} = \int_{S_3} \boldsymbol{j} \cdot d\boldsymbol{S} = 0, \qquad (11)$$

since no conduction current pierces $S_3$. This shows that Ampere's law cannot be used in the case of an open circuit including a capacitor. Maxwell's introduction of the displacement current density term fixed this difficulty. Applying now the integral form of (7) to $S_3$, one gets

$$\oint_{C_1} \boldsymbol{H} \cdot d\boldsymbol{r} = \int_{S_3} \frac{\partial \boldsymbol{D}}{\partial t} \cdot d\boldsymbol{S} \qquad (12)$$

Here we remember the law of charge conservation (continuity equation),

$$I + \frac{dQ}{dt} = \oint_S \boldsymbol{j} \cdot d\boldsymbol{S} + \int_V \frac{\partial \rho}{\partial t} dV = 0, \quad \nabla \cdot \boldsymbol{j} + \frac{\partial \rho}{\partial t} = 0 \qquad (13)$$

where S is the closed surface enclosing an arbitrary volume V, $Q$ is the total charge in the volume V and $I$ is the total flow of current out of the volume V. Then, with Gauss's law (4), the right hand side of (12) is

$$\int_{S_3} \frac{\partial \boldsymbol{D}}{\partial t} \cdot d\boldsymbol{S} = \frac{d}{dt} \int_{S_3} \boldsymbol{D} \cdot d\boldsymbol{S} = \frac{d}{dt} \oint_{\bar{S}_1 + S_3} \boldsymbol{D} \cdot d\boldsymbol{S} = \frac{dQ}{dt} = I \qquad (14)$$

where $\bar{S}_1 + S_3$ indicates the closed surface comprising of $\bar{S}_1$ and $S_3$, $\bar{S}_1$ being the same surface as $S_1$ but with the normal directed in the direction opposite to the case in (9) indicated by $d\boldsymbol{S}$ for $S_1$ in figure 1, and $Q$ is the total electric charge inside the volume bounded by $\bar{S}_1 + S_3$, i.e., that on the bottom electrode. The term after the second equal sign in equation (14) holds since we assume that $S_1$ is far enough away from the capacitor that $\boldsymbol{D} = 0$ on it and the value of the surface integral does not change when $\bar{S}_1$ is added to the integral area. Equations (12) and (14) show that the choice of $S_3$ gives the same result as

(9) using $S_1$ and $S_2$ owing to the displacement current density term.

Now let us consider the application of the law to calculate the magnetic field intensity $H_2$ at point $P_2$ in figure 2 using the surface $S_4$,

$$\oint_{C_2} \boldsymbol{H} \cdot d\boldsymbol{r} = \int_{S_4} \frac{\partial \boldsymbol{D}}{\partial t} \cdot d\boldsymbol{S} \approx I \qquad (15)$$

leading to

$$H_2 \approx \frac{I}{2\pi R} \qquad (16)$$

The approximation (actually $H_2 \leq H_1$) is due to a fraction of the fringing electric field not piercing $S_4$ [7, 10] and approaches equality as $R$ increases or as the distance between the electrodes decreases. This calculation may give the confusing first impression that the magnetic field there is created by $\partial \boldsymbol{D}/\partial t$ in the space between the electrodes.

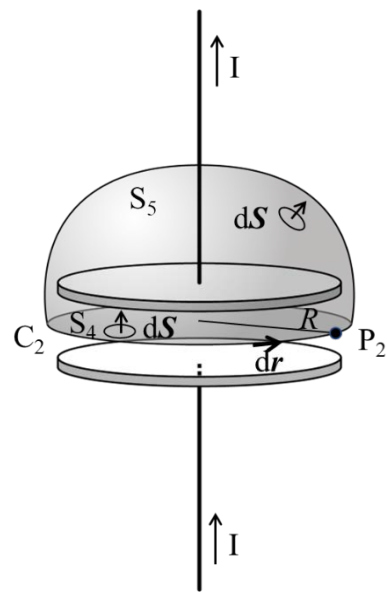

Figure 2. The same setup as figure 1. Point $P_2$ is on a circle $C_2$ of radius $R$, centered on the middle of the virtual straight line connecting the linear currents, and parallel to the capacitor electrodes.

Although it is certainly plausible to calculate the magnetic field by using (15), it does not mean that the impression is correct, for the same field can be calculated by using the surface $S_5$ to which the main contribution comes from the conduction current $I$ (rest is the partially cancelling contribution from the fringing $\partial \boldsymbol{D}/\partial t$). Mathematical plausibility of the calculation using the surface $S_4$ is warranted by the close correlation between the two fields, $\boldsymbol{B}$ and $\partial \boldsymbol{D}/\partial t$, both of which result from the same conduction currents flowing in the leads and the electrodes (see Sec. 3). Thus, both in figures 1 and 2, the *choice* of the surfaces in the calculation is irrelevant to whatever is the source of the field at point $P_1$ or $P_2$. Drawing students' attention to these points explicitly when first introducing the use of the integral form of the Ampere-Maxwell law will considerably reduce their misunderstanding.

## 3. Biot-Savart law and Ampere-Maxwell law

Another law that one may use to calculate magnetic field is the Biot-Savart law

$$\boldsymbol{H}(\boldsymbol{r}) = \int \frac{\boldsymbol{j}(\boldsymbol{r}') \times (\boldsymbol{r} - \boldsymbol{r}')}{4\pi |\boldsymbol{r} - \boldsymbol{r}'|^3} dV' \qquad (17)$$

Expressions for surface current density $\boldsymbol{J}_S(\boldsymbol{r}')$ flowing over a surface S or a current $I$ flowing in a thin wire may be obtained by replacing $\boldsymbol{j}(\boldsymbol{r}')dV'$ with $\boldsymbol{J}_S(\boldsymbol{r}')dS'$ or $Id\boldsymbol{r}'$ as

$$\boldsymbol{H}(\boldsymbol{r}) = \int_S \frac{\boldsymbol{J}_S(\boldsymbol{r}') \times (\boldsymbol{r} - \boldsymbol{r}')}{4\pi|\boldsymbol{r} - \boldsymbol{r}'|^3} dS', \qquad \boldsymbol{H}(\boldsymbol{r}) = \oint_C \frac{Id\boldsymbol{r}' \times (\boldsymbol{r} - \boldsymbol{r}')}{4\pi|\boldsymbol{r} - \boldsymbol{r}'|^3} \qquad (18)$$

respectively. This law describes how current density or a current element at $\boldsymbol{r}'$ contributes to the magnetic field at a point $\boldsymbol{r}$. (In Maxwell's equations as well, the expressions involving $\boldsymbol{j}$ and $\rho$ are only valid when they are finite. In the case of surface density and line intensity where $\boldsymbol{j}$ and $\rho$ are $\infty$, only the integral forms hold, applying the conversions shown above. This usual convention is used throughout the present paper.) This law is also applicable to an open circuit with a capacitor by adding up the contributions from all the conduction currents including, in the present case, the current spreading/centering radially over the electrode plates to make the charge on the plates increase/decrease [1, 2, 7, 9-12] as shown in figure 3. It is not only applicable to a static case but also approximately to a quasi-static case where the finiteness of the speed of light may be ignored [13, 14]. Furthermore, generalization of the Biot-Savart law to the time-dependent situation

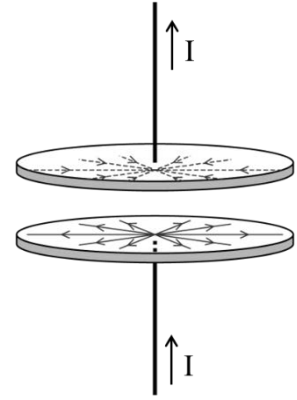

Figure 3. A parallel-plate capacitor being charged by a current I. Also shown are radial currents spreading out on the inside surface of the bottom plate, charging it positively, and those centering in on the inside surface of the top plate, charging it negatively.

$$\boldsymbol{B}(\boldsymbol{r},t) = \frac{\mu_0}{4\pi} \int \left[ \frac{\boldsymbol{j}(\boldsymbol{r}', t_r)}{R^2} + \frac{\partial \boldsymbol{j}(\boldsymbol{r}', t_r)/\partial t_r}{cR} \right] \times \frac{\boldsymbol{R}}{R} dV' \qquad (19)$$

has been developed by Jefimenko [15-20], where $\boldsymbol{R} = \boldsymbol{r} - \boldsymbol{r}'$, $c$ is the speed of light and

$$t_r = t - \frac{|\boldsymbol{r} - \boldsymbol{r}'|}{c} = t - \frac{R}{c} \qquad (20)$$

is the retarded time (former time). It reduces to the original Biot-Savart law (17) in the stationary case. Equations (17)-(19) show that the conduction currents are the only primary source of the magnetic field. These are certainly cause-effect type laws.

A question closely related with the present issue is whether displacement current density in the space between the electrodes should be included in the *current* density in (17) and (19). One answer is that, if its effect is directly calculated by the Biot-Savart law, it turns out to be 0 [1]; see also [18] and the references therein. More fundamentally the displacement current density related with the conservative or irrotational electric field whose curl is zero such as that between the capacitor electrodes does not result in a magnetic field [6, 7, 10-12].

Bartlett [11] made an analytical calculation of the magnetic field between the capacitor plates to show with some approximation that it is actually created by the linear current in the lead wire and the radial current in the plates. Milsom [12] provided numerical results together with an excellent compact review of the topic. Previously the calculation had only been explained in principle [10].

Next, let's return to check the integral form of the Ampere-Maxwell law. The differential form of Maxwell's equations describe the relationship between the time and space derivatives of the electric and magnetic fields and the charge and current densities at a given position and time, and are valid at any place and time. It is possible to regard them as simultaneous differential equations, express them with a scalar and vector potentials, and get solutions (retarded potentials) which have forms showing causality. However, the integral form of the equations are not the integral solutions of the differential form. The former is derived by applying vector analysis to the respective latter with time variable $t$ fixed throughout the whole space. For example, surface integration of both sides of the differential form of (6) on an arbitrary surface S bordered by any closed curve C yields

$$\int_S (\nabla \times \boldsymbol{H}(\boldsymbol{r},t)) \cdot d\boldsymbol{S} = \int_S \left(\boldsymbol{j}(\boldsymbol{r},t) + \frac{\partial \boldsymbol{D}(\boldsymbol{r},t)}{\partial t}\right) \cdot d\boldsymbol{S} \qquad (21)$$

Then rewriting the left-hand side as a line integral along C from Stokes' theorem, we obtain the integral form of (6). That is, the integral form represents the correlation or association between the integral of the field $\boldsymbol{H}$ along C and that of $\boldsymbol{j} + \partial \boldsymbol{D}/\partial t$ on S *at a common fixed time* in an inertial frame of reference. Thus the integral form of the Ampere-Maxwell law cannot be regarded as a cause-effect type law. Moreover it is only when the conduction current and the displacement current distributions are highly symmetric that we may calculate the magnetic field at a point by reducing the integral along the path to a product such as $2\pi R H_1 = I$, leading to (10). It should be remembered, on the other hand, that law itself holds in the above sense even in a very rapidly changing time-dependent field regardless of the size of the integration domain. A subtle consequence is that the integral form of the Ampere-Maxwell law which represents the relationships of the fields at a given time could be applied to calculate approximately the magnetic field in a quasi-static case.

## 4. Reported measurements of displacement current

There have been many papers published whose titles include "measurement of displacement current", "effect of displacement current" and so on, as cited by Roche [2] and French [9], the oldest of which is probably by Thompson [21]. More recent articles include reference [22]. All these experiments, and likely many other reports on this topic, take it for granted that the displacement current density, or time derivative of the electric field multiplied by $\varepsilon_0$, $\varepsilon_0 \boldsymbol{E}/\partial t$, in the space between the electrodes of a capacitor creates the magnetic field in and around it. Most of them measure in fact the electromotive force induced in toroidal search coils by the changing magnetic field in the space. Direct measurements of the magnetic field are also reported by Bartlet and coworkers [23, 24]. It is apparent that none of the reports measure the displacement current itself directly. Only if it were independently established that the displacement current is really causing the magnetic field in these situations, could they be considered to be at best indirect measurements of the displacement current. As we discussed above, however, the magnetic field is actually created by the conduction current only and merely has a correlation with the displacement current created at the same time.

5. **Two kinds of electric fields**

If the displacement current density between the capacitor electrodes does not create a magnetic field, one might ask why the displacement current density in the Ampere-Maxwell law is essential for the existence of electromagnetic waves. To answer this question it is crucial to note that the electric field in Maxwell's equations contains two kinds of electric fields [25].

One is directly responsible for Gauss's law (4), or using (2)

$$\oint_S \boldsymbol{E} \cdot d\boldsymbol{S} = \frac{1}{\varepsilon_0} \int_V \rho \, dV, \qquad \nabla \cdot \boldsymbol{E} = \frac{\rho}{\varepsilon_0} \qquad (22)$$

We denote this electric field by $\boldsymbol{E}_C$. As a matter of course it satisfies

$$\oint_S \boldsymbol{E}_C \cdot d\boldsymbol{S} = \frac{1}{\varepsilon_0} \int_V \rho \, dV, \qquad \nabla \cdot \boldsymbol{E}_C = \frac{\rho}{\varepsilon_0} \qquad (23)$$

This field is due to the electric charge, whose $r^{-2}$ dependence around a point charge leads to its conservative or irrotational nature

$$\oint_C \boldsymbol{E}_C \cdot d\boldsymbol{r} = 0, \qquad \nabla \times \boldsymbol{E}_C = 0 \qquad (24)$$

where, as before, C indicates arbitrary closed path in space.

The other electric field is directly responsible for Faraday's law (5). It is the electromagnetic induced field, and we denote it as $\boldsymbol{E}_I$. It satisfies

$$\oint_C \boldsymbol{E}_I \cdot d\boldsymbol{r} = -\frac{d}{dt} \int_S \boldsymbol{B} \cdot d\boldsymbol{S}, \qquad \nabla \times \boldsymbol{E}_I = -\frac{\partial \boldsymbol{B}}{\partial t} \qquad (25)$$

where S, as before, is any surface bordered by any closed path C in space. The left-hand side of the integral form of (25), and also that of (5), represents the electromotive force $\mathcal{E}$ along one turn of C;

$$\mathcal{E} = \oint_C \boldsymbol{E}_I \cdot d\boldsymbol{r} = -\frac{d}{dt}\int_S \boldsymbol{B} \cdot d\boldsymbol{S} \qquad (26)$$

Since $\mathcal{E} \neq 0$, $\boldsymbol{E}_I$ is not conservative, unlike $\boldsymbol{E}_C$ which satisfies (24).

Maxwell's equations express the electric field simply by $\boldsymbol{E}$ without distinguishing $\boldsymbol{E}_C$ and $\boldsymbol{E}_I$. Since these fields are different and mutually independent as shown above, this means that

$$\boldsymbol{E} = \boldsymbol{E}_C + \boldsymbol{E}_I \qquad (27)$$

It is not possible to observe the two fields $\boldsymbol{E}_C$ and $\boldsymbol{E}_I$ separately by measurement since they have exactly the same property of electric field on a charge as defined by (1). But we are sure that the electric field in the absence of changing magnetic field is $\boldsymbol{E}_C$ while that in the absence of electric charge is $\boldsymbol{E}_I$.

In the following, we will examine the relationship between the electric field $\boldsymbol{E}$ appearing explicitly in Maxwell's equations and the two fields $\boldsymbol{E}_C$ and $\boldsymbol{E}_I$ that comprise it, and then point out that overlooking the particular property of $\boldsymbol{E}$ could lead to misunderstandings about the nature of the displacement current density addressed in this paper.

First we check Faraday's law of magnetic induction. Substituting (27) into (5) yields

$$\oint_C (\boldsymbol{E}_C + \boldsymbol{E}_I) \cdot d\boldsymbol{r} = -\frac{d}{dt}\int_S \boldsymbol{B} \cdot d\boldsymbol{S}, \quad \nabla \times (\boldsymbol{E}_C + \boldsymbol{E}_I) = -\frac{\partial \boldsymbol{B}}{\partial t} \qquad (28)$$

By subtracting (25) from this, we obtain

$$\oint_C \boldsymbol{E}_C \cdot d\boldsymbol{r} = 0, \quad \nabla \times \boldsymbol{E}_C = 0 \qquad (29)$$

which is the same as (24). The conservative or irrotational nature of the $\boldsymbol{E}_C$, which is originally derived from the $r^{-2}$ dependency of the field, is derived from (5), (25) and (27).

Because of (29) the law for $\boldsymbol{E}_I$, (25), holds for $\boldsymbol{E}$ including $\boldsymbol{E}_C$ as (5) without affecting the mathematical correctness of the equation. In other words, even though $\boldsymbol{E}_C$ is included in the electric field $\boldsymbol{E}$ in equation (5), it is irrelevant to Faraday's law of magnetic induction.

Next let's check Gauss' law (4) with (2), i.e., (22). Substituting (27) into (22) yields

$$\oint_S (\boldsymbol{E}_C + \boldsymbol{E}_I) \cdot d\boldsymbol{S} = \frac{1}{\varepsilon_0}\int_V \rho \, dV, \quad \nabla \cdot (\boldsymbol{E}_C + \boldsymbol{E}_I) = \frac{\rho}{\varepsilon_0} \qquad (30)$$

By subtracting (23) from this, we obtain

$$\oint_S \boldsymbol{E}_I \cdot d\boldsymbol{S} = 0, \quad \nabla \cdot \boldsymbol{E}_I = 0 \qquad (31)$$

This means that $\boldsymbol{E}_I$ is divergence-free everywhere, unlike $\boldsymbol{E}_C$ satisfying (24). In other words, even though $E_I$ is included in the electric field $E$ in equation (4), it is irrelevant to Gauss' law. Note that equation (31) cannot be derived independently from the nature of $E_I$, but is deduced from within Maxwell's equations. Helmholtz theorem guarantees that any vector field that goes to zero faster than $1/r$ as $r \to \infty$ can be decomposed into irrotational and divergence-free fields [26]. It now turns out that (27) represents such a decomposition of the electric field $E$ in Maxwell's equations, which itself is neither irrotational nor divergence-free.

Next, let's check the Ampere-Maxwell law (7). Taking the divergence of both sides of its differential form after using (2) and (3), we get

$$\nabla \cdot (\nabla \times B) = \mu_0 \nabla \cdot j + \varepsilon_0 \mu_0 \frac{\partial}{\partial t} \nabla \cdot E \qquad (32)$$

The left-hand side is zero, and substituting (27) for the second term on the right-hand side, we obtain

$$\nabla \cdot j + \varepsilon_0 \frac{\partial}{\partial t} \nabla \cdot (E_C + E_I) = 0 \qquad (33)$$

and taking (31) into account,

$$\nabla \cdot j + \nabla \cdot \left(\varepsilon_0 \frac{\partial E_C}{\partial t}\right) = 0 \qquad (34)$$

Referring to (23), it is evident that (34) reduces to the law of charge conservation (13). Equation (34) shows that $\varepsilon_0 \partial E_C/\partial t$ is the sole displacement current density effectively involved in (13). In other words, <u>even though $\varepsilon_0 \partial E_I/\partial t$ is included in the displacement current density term $\partial D/\partial t$ in (7), it is irrelevant to the law of charge conservation.</u>

Finally let us consider the derivation of the differential equations for electromagnetic waves. If we take the curl of both sides of the differential form of (5), we have

$$\nabla \times \nabla \times E = -\frac{\partial}{\partial t} \nabla \times B \qquad (35)$$

Substituting (27) into the left-hand side of (35) yields

$$\nabla \times \nabla \times (E_C + E_I) = \nabla \times \nabla \times E_I = \nabla(\nabla \cdot E_I) - \nabla^2 E_I = -\nabla^2 E_I \qquad (36)$$

We have used first (24) and then the vector analysis identity and (31). From (7) with (2) and (3), the right-hand side of (35) becomes

$$-\frac{\partial}{\partial t} \nabla \times B = -\mu_0 \frac{\partial j}{\partial t} - \varepsilon_0 \mu_0 \frac{\partial}{\partial t}\left(\frac{\partial E_C}{\partial t} + \frac{\partial E_I}{\partial t}\right) \qquad (37)$$

Thus we have

$$\nabla^2 E_I = \mu_0 \frac{\partial j}{\partial t} + \varepsilon_0 \mu_0 \frac{\partial}{\partial t}\left(\frac{\partial E_C}{\partial t} + \frac{\partial E_I}{\partial t}\right) \qquad (38)$$

Now remembering Helmholtz theorem, we decompose the current density into the sum of the irrotational (often called longitudinal) part $\boldsymbol{j}_L$ and the divergence-free (often called transverse) part $\boldsymbol{j}_T$

$$\boldsymbol{j} = \boldsymbol{j}_L + \boldsymbol{j}_T \qquad (39)$$

where

$$\nabla \times \boldsymbol{j}_L = 0 \quad \text{and} \quad \nabla \cdot \boldsymbol{j}_T = 0 \qquad (40)$$

Then (38) is separated into two:

$$\nabla^2 \boldsymbol{E}_I - \varepsilon_0 \mu_0 \frac{\partial^2 \boldsymbol{E}_I}{\partial t^2} = \mu_0 \frac{\partial \boldsymbol{j}_T}{\partial t} \qquad (41)$$

$$\frac{\partial}{\partial t}\left(\boldsymbol{j}_L + \varepsilon_0 \frac{\partial \boldsymbol{E}_C}{\partial t}\right) = 0 \qquad (42)$$

Equation (41) is the inhomogeneous differential equation for the electric field part of the electromagnetic wave. It shows that $\varepsilon_0 \partial \boldsymbol{E}_I/\partial t$ is the sole displacement current density effectively involved in the derivation of the electromagnetic waves. In other words, <u>even though $\varepsilon_0 \partial \boldsymbol{E}_C/\partial t$ which is related to the conservative electric field as that between the electrodes of a capacitor is included in $\partial \boldsymbol{D}/\partial t$ in (7), it is irrelevant to the electromagnetic waves.</u> Taking the divergence of (42) and using (23), we have

$$\frac{\partial}{\partial t}\left(\nabla \cdot \boldsymbol{j}_L + \frac{\partial}{\partial t}\rho\right) = 0 \qquad (43)$$

Noting that $\nabla \cdot \boldsymbol{j} = \nabla \cdot \boldsymbol{j}_L$, this is the time derivative of the law of charge conservation (13), showing that (13) always holds.

For the magnetic field part of the wave equation, we take the curl of (7) with (2), (3), (27) and (39), obtaining

$$\nabla \times \nabla \times \boldsymbol{B} = \mu_0 \nabla \times (\boldsymbol{j}_L + \boldsymbol{j}_T) + \varepsilon_0 \mu_0 \nabla \times \left(\frac{\partial \boldsymbol{E}_C}{\partial t} + \frac{\partial \boldsymbol{E}_I}{\partial t}\right) = \mu_0 \nabla \times \boldsymbol{j}_T + \varepsilon_0 \mu_0 \nabla \times \frac{\partial \boldsymbol{E}_I}{\partial t} \qquad (44)$$

Taking into account (6) in the left-hand side after using the vector analysis identity as in (36), substituting (25) into the right-hand side, and then collecting the terms containing $\boldsymbol{B}$ on the left-hand side, we have the inhomogeneous wave equation for the magnetic field part of the electromagnetic wave

$$\nabla^2 \boldsymbol{B} - \varepsilon_0 \mu_0 \frac{\partial^2 \boldsymbol{B}}{\partial t^2} = -\mu_0 \nabla \times \boldsymbol{j}_T \qquad (45)$$

Equation (44) shows, again, that $\varepsilon_0 \partial \boldsymbol{E}_I/\partial t$ is the sole electric displacement current density effectively involved in the derivation of the electromagnetic waves.

**Summary**

Possible causes of the misconception about the nature of the displacement current density in Maxwell's equations are discussed. Firstly, an example of applying the integral form of the Ampere-Maxwell law to calculate the magnetic field in and around a parallel-plate capacitor by using a plane between the electrodes and parallel to the electrodes may lead to the misunderstanding that the displacement current density between the capacitor electrodes could be a source of the magnetic field. It is urged that students' attention is drawn to the fact that the choice of surface used in the calculation does not, of course, determine the source of the magnetic field obtained. Secondly, misinterpretation of the integral form of the Ampere-Maxwell law as expressing a cause-effect relationship could cause confusion. It is in fact a formula showing a correlation or association of the fields. Thirdly, it has been pointed out that the electric field $E$ appearing in Maxwell's equations is the sum of the conservative field $E_C$ and the induction field $E_I$. How they are involved in Maxwell's equations has been carefully explored with the aim of identifying the underlying cause of the controversies concerning the displacement current. It is shown, in particular, that only the displacement current density related with $E_C$ makes Maxwell's equations consistent with the law of charge conservation, while only that related with $E_I$ plays a crucial role in the derivation of the differential equations for electromagnetic waves. Furthermore, additional support provided from the calculations using the Biot-Savart law, which show that the magnetic field between the capacitor plate is actually created by the real currents alone, have only recently been reported. This late confirmation may have been another factor which allowed the misconception to persist for a long time.

**Acknowledgement**

The author thanks Dr. Shozo Suto, Dr. Satoshi N. Nakamura and Dr. Nazrene Zafar for valuable discussions.**ORCID iDs**

Toshio Hyodo: https://orcid.org/0000-0002-1810-732X